\documentclass[6p]{elsarticle}

\usepackage{amssymb}
\usepackage{graphicx}
\usepackage{xcolor}
\usepackage{amsmath}
\usepackage{lineno}
\usepackage{url}
\usepackage{multirow}
\usepackage{booktabs}
\journal{Journal Name}
\usepackage{float}
\begin{document}

\begin{frontmatter}

\title{Influence of Coherent Elastic Strain on Phase Separation in BCC Nb--V Alloys}

\author[tamu]{Siya Zhu\corref{cor1}}
\author[tamu,tamu_meen,tamu_msen]{Raymundo Arr\'{o}yave}
\cortext[cor1]{Corresponding author\\ Email: siyazhu@tamu.edu}

\address[tamu]{Department of Materials Science and Engineering, Texas A\&M University, College Station, TX 77843}
\address[tamu_meen]{J. Mike Walker '66 Department of Mechanical Engineering, Texas A\&M University, College Station, TX 77843}
\address[tamu_msen]{Wm Michael Barnes '64 Department of Industrial and Systems Engineering, Texas A\&M University, College Station, TX 77843}

\begin{abstract}
    Coherent elastic strain is an important but often neglected contribution to phase-separation thermodynamics in alloy systems where decomposed phases have appreciable lattice mismatch. We develop a thermodynamic framework that incorporates coherent elastic compatibility directly into phase-diagram calculations alongside conventional CALPHAD chemical free energies. Applied to the BCC Nb--V system, the framework shows that coherent elasticity substantially suppresses phase separation, narrows the miscibility gap, and lowers the critical temperature toward experimentally observed values. Beyond these quantitative effects, the coherent constraint qualitatively alters the interpretation of phase equilibria: the equilibrium decomposition compositions become functions of both temperature and overall alloy composition, so the two-phase boundary no longer represents unique coexistence compositions. These results establish coherent elasticity as a key thermodynamic factor in lattice-mismatched systems and provide a general framework for coherent phase-diagram modeling.
\end{abstract}
\end{frontmatter}

\section{Introduction}

Equilibrium phase boundaries in alloys are conventionally described within the CALPHAD framework, where phase stability is determined by the chemical free energies of stress-free phases. In practice, however, decomposed phases often remain at least partially coherent, so lattice compatibility across the interface introduces elastic strain energy in addition to the chemical driving force. This contribution is especially important in systems with large lattice mismatch, where the strain energy can become comparable to the chemical driving force itself. Coherent elasticity may therefore alter both the phase fractions and equilibrium compositions of coexisting phases, yet this effect is rarely incorporated explicitly into miscibility-gap calculations.

The BCC niobium--vanadium binary alloy provides an ideal model system for examining this problem. Nb and V share the same BCC crystal structure and exhibit complete high-temperature solubility, yet possess a substantial size mismatch, with lattice parameters of approximately 3.30 \AA{} for pure Nb and 3.02 \AA{} for pure V at ambient conditions. This places Nb--V in a thermodynamic regime where coherent elastic effects can compete directly with the chemical tendency toward phase separation.

Early experimental assessments described Nb--V as a continuous BCC solid solution without any confirmed low-temperature transformation~\cite{rudy1969compendium, wilhelm1954columbium, balakrishna1979synthesis, smith1983nb}. Gao \emph{et al.}~\cite{gao2018experimental} later pointed out that atomic diffusion in refractory Nb--V alloys is extremely sluggish at low temperatures, so the relatively short annealing times used in earlier studies were insufficient to rule out a miscibility gap. By extending annealing treatments to 1680 h, together with diffusion-couple experiments and differential thermal analysis, they reported clear evidence for a miscibility gap below 1077~K.

Several theoretical studies have also predicted a miscibility gap in Nb--V. Zhu \emph{et al.}~\cite{zhu2025accelerating} used Special Quasirandom Structures (SQS)~\cite{zunger1990special} combined with first-principles calculations and machine-learning interatomic potentials to construct a CALPHAD model, yielding a critical temperature of approximately 1400~K. Ravi \emph{et al.}~\cite{ravi2012first} combined density-functional theory (DFT), cluster expansion (CE), and Monte Carlo (MC) simulations, predicting a critical temperature of approximately 1250~K from a conventional CE description that decreased to about 950~K after including constituent strain and phonon contributions. Wang \emph{et al.}~\cite{wang2023effective} reported a critical temperature of approximately 1280~K from a chemical-only CE model, which decreased to around 1100~K after including lattice distortion effects. Kumar and Jindal~\cite{kumar2022first} incorporated short-range-order (SRO) effects through Cluster Variation Methods (CVM) and reassessed the Nb--V system using the experimental data of Gao \emph{et al.} Despite these corrections, theoretical studies consistently predict a broader miscibility gap than observed experimentally, suggesting that additional stabilizing factors remain unaccounted for in current models.

We propose that coherent elastic strain between decomposed phases is one such factor. In this work, we develop a thermodynamic framework that combines composition-dependent first-principles chemical free energies with explicit coherent elastic contributions, and apply it to the Nb--V system. Two limiting descriptions are examined: an isotropic common-volume model and a biaxial coherent model that allows out-of-plane relaxation while preserving in-plane lattice matching. We show that coherent elastic strain substantially reduces the predicted miscibility gap and lowers the critical temperature toward experimental values. Under coherent constraints, the equilibrium compositions of the two decomposed phases depend on the overall alloy composition, so the conventional common-tangent tie-line picture no longer strictly applies.

\section{Computational Methods}

\subsection{Thermodynamic model}

The Gibbs free energy of a single phase at composition $x$, temperature $T$, and strain $\epsilon$ is written as the sum of a chemical and an elastic contribution,
\begin{equation}
\begin{aligned}
    G(x,T,\epsilon) &= G_{\mathrm{chem}}(x,T) + G_{\mathrm{el}}(x,T,\epsilon)\\
    &= H(x) - TS(x) + G_{\mathrm{el}}(x,T,\epsilon)\\
    &= E_0(x) + PV(x) - T\left[S_\mathrm{conf}(x)+S_\mathrm{other}(x)\right]+G_{\mathrm{el}}(x,T,\epsilon)\\
    &\approx E_0(x) + k_\mathrm{B}T \left[x\ln x + (1-x)\ln(1-x)\right]+G_{\mathrm{el}}(x,T,\epsilon)
\end{aligned}
\end{equation}
Here, \(G_{\mathrm{chem}}\) is the composition-dependent chemical free energy, including the 0~K reference energy and configurational entropy, and \(G_{\mathrm{el}}\) is the elastic strain energy arising from the coherent lattice constraint. The \(PV\) term is neglected for condensed metallic phases, the electronic entropy is omitted, and the vibrational entropy is approximated as linear in composition so that it cancels in free-energy comparisons between decomposed states. The configurational entropy is described by the ideal random-solution expression.

For a homogeneous single-phase alloy at overall composition \(X\) and temperature \(T\), the free energy is denoted
\begin{equation}
G^{1\phi}(X,T)
\end{equation}

For a two-phase state with phase compositions \(x_1\) and \(x_2\) ($x_1<X<x_2$), the total free energy is
\begin{equation}
G^{2\phi}(X,T,\epsilon)
=
\frac{x_2-X}{x_2-x_1}G(x_1,T,\epsilon) + \frac{X-x_1}{x_2-x_1}G(x_2,T,\epsilon),
\end{equation}
where each phase carries both chemical and elastic contributions under the coherent constraint. The strain variable $\epsilon$ represents the structural degrees of freedom associated with coherent coexistence and is determined by minimizing the total free energy subject to mechanical compatibility constraints.

The equilibrium phase state at each \((X, T)\) is determined by comparing the single-phase free energy with the minimum over all admissible coherent two-phase states. The two-phase state is favored when
\begin{equation}
\min_{x_1,x_2,\epsilon}
G^{2\phi}(X,T,x_1,x_2,\epsilon)
<
G^{1\phi}(X,T).
\end{equation}
The boundary obtained from this comparison is the coherent binodal.

We also evaluate the coherent spinodal, which marks the local instability of the homogeneous phase against infinitesimal composition fluctuations. In chemical-only models, the spinodal is identified from the curvature condition
\begin{equation}
\frac{\partial^2 G}{\partial x^2} < 0,
\end{equation}
In the present framework, however, the free energy depends on additional strain variables determined through constrained minimization, so the Hessian cannot be obtained in closed form.

Instead, the coherent spinodal is determined numerically. The homogeneous state at composition \(X\) is spinodally unstable when, for a small perturbation \(\delta\),
\begin{equation}
\min_{\epsilon}\left[G(X-\delta,T,\epsilon)+G(X+\delta,T,\epsilon)\right]-2G(X,T)<0,
\end{equation}
where the free energy at each perturbed composition is re-optimized with respect to the coherent strain variables.

\subsection{Isotropic coherent elastic model}
To evaluate the coherent elastic contribution \(G_{\mathrm{el}}\), two mechanical constraint models are considered. The first is an isotropic coherent model based on the Birch--Murnaghan equation of state~\cite{murnaghan1944compressibility, birch1947finite}, in which the two coexisting phases share a common atomic volume. This choice is motivated by experimental reports describing phase-separated Nb--V alloys as two BCC phases with nearly identical lattice parameters.

In the isotropic model, a uniform strain is applied equally along all three crystallographic directions, with deformation gradient tensor
\begin{equation}
\mathbf F=
\begin{pmatrix}
\lambda & 0 & 0\\
0 & \lambda & 0\\
0 & 0 & \lambda
\end{pmatrix}
\end{equation}
where \(\lambda\) is the isotropic stretch ratio and is related to the engineering strain through
\begin{equation}
\lambda = 1+\epsilon
\end{equation}
The corresponding volume change is
\begin{equation}
\frac{V}{V_0}=\det(\mathbf F)=(1+\epsilon)^3.
\end{equation}
where \(V_0\) is the equilibrium volume at zero pressure. To describe the finite volumetric deformation, we employ the conventional Eulerian (Euler--Almansi) strain tensor,

\begin{equation}
\mathbf e^{\mathrm{EA}}
=
\frac12\left(\mathbf I-\mathbf F^{-\mathrm T}\mathbf F^{-1}\right),
\end{equation}

which reduces, for the present isotropic deformation, to the scalar Eulerian strain as
\begin{equation}
\eta\equiv-e^{\mathrm{EA}} =\frac12\left[(1+\epsilon)^{-2}-1\right]=\frac12\left[\left(\frac{V_0}{V}\right)^{2/3}-1\right]
\end{equation}
The strain energy is expanded to the third-order as
\begin{equation}
E(\eta)=E_0 + A_2 \eta^2 + A_3 \eta^3 + O(\eta^4)
\label{eq:iso_expansion}
\end{equation}
where \(E_0\) is the equilibrium energy, and \(A_2\) and \(A_3\) are expansion coefficients. The pressure $P$ is obtained from
\begin{equation}
    \begin{aligned}
         P & = -\frac{dE}{dV}
            = -\frac{dE}{d\eta}\frac{d\eta}{dV} \\
           & = \frac{1}{3V}\left(\frac{V_0}{V}\right)^{2/3}   \left[2A_2\eta+3A_3\eta^2+O(\eta^3)\right]\\
           & = \frac{1}{3V_0}\left(1+2\eta\right)^{5/2}\left[2A_2\eta+3A_3\eta^2+O(\eta^3)\right]\\
           & = \frac{1}{3V_0}\left[2A_2\eta + \left(10A_2+3A_3\right)\eta^2\right]+O(\eta^3)
    \end{aligned}
\end{equation}
The bulk modulus $B$ is defined as
\begin{equation}
    \begin{aligned}
         B & = -V\frac{dP}{dV}
             = -V\frac{dP}{d\eta}\frac{d\eta}{dV}\\
           & = -V\cdot\frac{2A_2+2
           \left(10A_2+3A_3\right)\eta+O(\eta^2)}{3V_0}\cdot\frac{-\left(1+2\eta\right)^{5/2}}{3V_0} \\
           & = \frac{2A_2}{9V_0}+\frac{8A_2+2A_3}{3V_0}\eta+O(\eta^2)
    \end{aligned}
\end{equation}
The equilibrium bulk modulus $B_0$ at $\eta=0$ is 
\begin{equation}
    \begin{aligned}
        B_0 & = \frac{2A_2}{9V_0}
    \end{aligned}
\end{equation}
and the pressure derivative of $B_0$ is
\begin{equation}
    \begin{aligned}
        B_0' & = \left.\frac{dB}{dP}\right|_{\eta=0}
              = \left.\frac{dB}{d\eta}\frac{d\eta}{dP}\right|_{\eta=0}\\
             & = \frac{8A_2+2A_3}{3V_0}\cdot\frac{3V_0}{2A_2}\\
             & = 4 + \frac{A_3}{A_2}
    \end{aligned}
\end{equation}

The expansion coefficients can therefore be rewritten in terms of the equilibrium bulk modulus $B_0$ and its pressure derivative $B_0'$:
\begin{equation}
A_2=\frac{9}{2}V_0B_0,
\qquad
A_3=\frac{9}{2}V_0B_0(B_0'-4).
\end{equation}
Substituting into Eq.~\ref{eq:iso_expansion} gives
\begin{equation}
    \begin{aligned}
        E & = E_0 + \frac{9}{2}V_0B_0 \eta^2 + \left[\frac{9}{2}V_0B_0(B_0'-4)\right] \eta^3\\
              & =  E_0 + \frac{9 V_0 B_0}{16}
                \Bigg\{ \left[
                \left(\frac{V_0}{V}\right)^{2/3} - 1
                \right]^3 B_0' \\
              & + \left[\left(\frac{V_0}{V}\right)^{2/3} - 1\right]^2\left[6 - 4\left(\frac{V_0}{V}\right)^{2/3}\right]\Bigg\}
    \end{aligned} 
\end{equation}

\subsection{Biaxial coherent elastic model}
We next introduce a biaxial coherent model to capture a more general interfacial constraint. In this case, the two phases are constrained to share the lattice parameter parallel to the coherent interface, while the lattice parameter perpendicular to the interface is fully relaxed. 

We describe the biaxial coherent deformation using the deformation gradient tensor

\begin{equation}
\mathbf F =
\begin{pmatrix}
\lambda_{\perp} & 0 & 0 \\
0 & \lambda_{\parallel} & 0 \\
0 & 0 & \lambda_{\parallel}
\end{pmatrix},
\end{equation}

where $\lambda_{\perp}$ and $\lambda_{\parallel}$ denote the stretch ratios along the out-of-plane and in-plane directions, respectively, and are related to the engineering strains through
\begin{equation}
\lambda_{\perp} = 1+\epsilon_{\perp},
\qquad
\lambda_{\parallel} = 1+\epsilon_{\parallel}
\end{equation}

To account for finite deformation effects, we employ the Green--Lagrange strain tensor,

\begin{equation}
\mathbf{E}^{\mathrm{GL}}
=
\frac{1}{2}\left(\mathbf F^{\mathrm{T}}\mathbf F-\mathbf I\right).
\end{equation}

For the present biaxial deformation, the two independent principal Green--Lagrange strain components are denoted by \(f_{\perp}\) and \(f_{\parallel}\), corresponding to the out-of-plane and in-plane directions, respectively:

\begin{equation}
f_{\perp}
\equiv
E_{\perp}^{\mathrm{GL}}
=
\frac{1}{2}\left(\lambda_{\perp}^{2}-1\right),
\qquad
f_{\parallel}
\equiv
E_{\parallel}^{\mathrm{GL}}
=
\frac{1}{2}\left(\lambda_{\parallel}^{2}-1\right).
\end{equation}

The elastic energy surface is then expressed as a third-order polynomial expansion in the finite-strain variables,
\begin{equation}
\begin{aligned}
E(f_{\perp}, f_{\parallel}) =\;& E_0
+ A_{20} f_{\perp}^{2}
+ A_{11} f_{\perp}f_{\parallel}
+ A_{02} f_{\parallel}^{2}
\nonumber\\
&+ A_{30} f_{\perp}^{3}
+ A_{21} f_{\perp}^{2}f_{\parallel}
+ A_{12} f_{\perp}f_{\parallel}^{2}
+ A_{03} f_{\parallel}^{3}.
\end{aligned}
\label{eq:biaxialexpand}
\end{equation}
The quadratic terms ($A_{20}, A_{02}$) describe the harmonic elastic response, the cubic terms ($A_{30}, A_{03}$) capture anharmonicity and tension--compression asymmetry, and the mixed terms ($A_{11}, A_{21}, A_{12}$) couple the in-plane and out-of-plane deformations.

Both Green--Lagrange and Eulerian finite-strain parameterizations were tested for the biaxial energy-surface fitting. The Green--Lagrange form yields a lower overall reconstruction error after composition interpolation and is therefore adopted here; details are provided in the Supplementary Information (Section~S3 and Fig.~S2-S4). For the isotropic model, the conventional Eulerian representation is retained, consistent with the standard Birch--Murnaghan formalism.

\subsection{Energy--strain datasets and composition-dependent fitting}

Composition-dependent elastic energy models were constructed from first-principles calculations on SQS~\cite{zunger1990special} generated with the ATAT package~\cite{van2002alloy, van2013efficient, van2017software} to represent chemically disordered BCC solid solutions at selected compositions. Each SQS structure was first fully relaxed to obtain the zero-stress reference state. Strained supercells were then generated by imposing prescribed lattice deformations; for each, the lattice vectors were held fixed while internal atomic coordinates were relaxed before a final static total-energy calculation. This procedure yields a consistent set of energies as a function of strain for fitting both the isotropic and biaxial elastic models.

For the isotropic coherent model, we used the optimized BCC SQS structures distributed with ATAT. Compositions were sampled from $x_{\mathrm{V}}=0,\,0.125,\,...,\,1.0$. For the pure end members ($x_{\mathrm{V}}=0$ and $1$), the one-atom primitive BCC cell was used. At $x_{\mathrm{V}}=0.5$, a 48-atom SQS was adopted, while all remaining intermediate compositions employed 32-atom SQS cells. For each relaxed reference structure, isotropically scaled cells were generated with relative volumes $V/V_0 = 0.90,\,0.92,\,...,\,1.10$, corresponding to eleven volumetric strain states. Each structure was then relaxed with fixed cell shape and volume, followed by a single-point energy calculation. These datasets were used to fit the composition-dependent Birch--Murnaghan parameters $E_0$, $V_0$, $B_0$, and $B_0'$.

For the biaxial coherent model, the pure Nb and V limits again used the one-atom primitive BCC cell. For intermediate compositions $x_{\mathrm{V}}=1/9,\,2/9,\,...,\,8/9$, we generated $3\times3\times3$ BCC SQS supercells containing 54 atoms. Starting from the relaxed equilibrium structure, a normal strain was first imposed along the $[100]$ direction with stretch ratios ranging from $0.90$ to $1.10$ in increments of $0.02$ (eleven states). For each imposed $[100]$ strain, an additional equibiaxial in-plane strain was applied simultaneously along the $[010]$ and $[001]$ directions, corresponding to the coherent $(100)$ interface plane, again using stretch ratios from $0.90$ to $1.10$. This generated an $11\times11$ grid of biaxial strain states, i.e., 121 strained configurations for each composition. For every strained cell, the lattice vectors were fixed while all internal atomic positions were relaxed prior to the final total-energy calculation. The resulting two-dimensional energy surfaces were then used to fit the third-order finite-strain coefficients of Eq.~\ref{eq:biaxialexpand}. 

All model coefficients --- \(E_0\), \(V_0\), \(B_0\), and \(B_0^{\prime}\) for the isotropic model, and \(E_0\) and \(A_{ij}\) for the biaxial model --- were fitted in two steps. First, parameters were independently determined at each sampled composition from the energy datasets described above. Second, the resulting values were represented as continuous functions of composition using a second-order Redlich--Kister (RK) expansion~\cite{redlich1948algebraic},
\begin{equation}
P(x_{\mathrm{V}})=
(1-x_{\mathrm{V}})P^{\mathrm{Nb}}
+x_{\mathrm{V}}P^{\mathrm{V}}
+x_{\mathrm{V}}(1-x_{\mathrm{V}})
\sum_{n=0}^{2} L_n(2x_{\mathrm{V}}-1)^n,
\end{equation}
where \(P(x_{\mathrm{V}})\) denotes the value of a given coefficient at composition \(x_{\mathrm{V}}\), \(P^{\mathrm{Nb}}\) and \(P^{\mathrm{V}}\) are the corresponding end-member values for pure Nb and pure V, respectively, and \(L_n\) are the RK interaction parameters. Both the end-member values and RK coefficients were treated as fitting parameters. This procedure provides continuous composition-dependent thermodynamic and elastic coefficients for subsequent free-energy minimization and phase-stability calculations.

To incorporate finite-temperature effects, only \(V_0\) was treated as temperature dependent; all other coefficients were held fixed. The thermal expansion coefficient was interpolated linearly between the pure elements,
\begin{equation}
\alpha(x_{\mathrm{V}})=(1-x_{\mathrm{V}})\alpha_{\mathrm{Nb}}+x_{\mathrm{V}}\alpha_{\mathrm{V}},
\end{equation}
where \(\alpha_{\mathrm{Nb}}\) and \(\alpha_{\mathrm{V}}\) are the linear thermal expansion coefficients of pure Nb and V. The finite-temperature equilibrium volume is then
\begin{equation}
V_0(x_{\mathrm{V}},T)=V_0(x_{\mathrm{V}},0)\exp\!\left(\int_0^T 3\alpha(x_{\mathrm{V}})\, dT \right).
\end{equation}
For constant \(\alpha(x_{\mathrm{V}})\), this reduces to
\begin{equation}
V_0(x_{\mathrm{V}},T)=V_0(x_{\mathrm{V}},0)\exp\!\left[3\alpha(x_{\mathrm{V}})T\right].
\end{equation}
and for small thermal strains the first-order approximation
\begin{equation}
V_0(x_{\mathrm{V}},T)\approx V_0(x_{\mathrm{V}},0)\left[1+3\alpha(x_{\mathrm{V}})T\right]
\end{equation}
is used in the present work.

\subsection{First-principles calculations}

All DFT calculations were performed using the Vienna Ab-initio Simulation Package (VASP)~\cite{kresse1993ab, kresse1994ab, kresse1996efficiency, kresse1996efficient} with the Perdew--Burke--Ernzerhof (PBE)~\cite{perdew1996generalized} exchange-correlation functional and PAW pseudopotentials~\cite{blochl1994projector}. The plane-wave cutoff energy was set to 600~eV, and Brillouin-zone integrations used a Monkhorst--Pack mesh~\cite{monkhorst1976special} with 8000 k-points per reciprocal atom.

\section{Results and Discussions}
\subsection{Composition-dependent isotropic equation-of-state parameters}

The isotropic energetic model described in Sec.~2.2 was constructed by first fitting the DFT energy--volume datasets at each sampled composition to the Birch--Murnaghan equation of state, yielding discrete values of the equilibrium energy \(E_0\), equilibrium atomic volume \(V_0\), bulk modulus \(B_0\), and its pressure derivative \(B_0^{\prime}\). These composition-dependent datasets were then further represented using the second-order RK expansion introduced in Sec~2.4, providing a continuous thermodynamic description across the full Nb--V binary range.

The resulting RK representation reproduces the calculated datasets with excellent accuracy. For \(E_0\), \(V_0\), and \(B_0\), near-unity coefficients of determination were obtained, indicating that their composition dependence is smooth and well captured by a second-order RK formalism. Although the fit quality for \(B_0^{\prime}\) is slightly lower than for the other quantities, the agreement remains satisfactory for subsequent phase-stability calculations. These results confirm that the present interpolation scheme provides a robust energetic model suitable for thermodynamic analysis.

The fitted end-member coefficients recovered from the RK model are summarized in Table~\ref{tab:endmember_compare}, together with representative experimental values. Good agreement is obtained for both Nb and V. In particular, the predicted equilibrium volumes closely reproduce the known lattice constants of BCC Nb and V, while the calculated bulk moduli and pressure derivatives remain within the expected experimental ranges.

\begin{table}[h]
\centering
\caption{End-member properties obtained from the second-order Redlich--Kister representation, compared with representative literature values for pure BCC Nb and V.}
\label{tab:endmember_compare}
\begin{tabular}{llccc}
\toprule
Phase & Property & Value & Source \\
\midrule
\multirow{8}{*}{Nb (bcc)}
& \(E_0\) (eV/atom)      & -10.2160 & Present work \\
& \(V_0\) (\AA$^3$/atom) & 18.0930  & Present work \\
&                         & 17.95    & Experiment~\cite{kittel2018introduction} \\
&                         & 17.98    & Experiment~\cite{kenichi2006high} \\
&                         & 18.32    & DFT~\cite{kovci2008elasticity} \\

& \(B_0\) (GPa)          & 170.55   & Present work \\
&                         & 170.69   & Experiment~\cite{katahara1979pressure} \\
&                         & 168   & Experiment~\cite{kenichi2006high} \\
&                         & 174   & DFT~\cite{kovci2008elasticity} \\
& \(dB_0/dP\)            & 3.7917   & Present work \\
&                         & 3.98     & Experiment~\cite{katahara1979pressure} \\
&                         & 3.4     & Experiment~\cite{kenichi2006high} \\
&                         & 3.98     & DFT~\cite{kovci2008elasticity} \\
\midrule
\multirow{8}{*}{V (bcc)}
& \(E_0\) (eV/atom)      & -8.9416  & Present work \\
& \(V_0\) (\AA$^3$/atom) & 13.2130  & Present work \\
&                         & 13.83    & Experiment~\cite{kittel2018introduction} \\
&                         & 13.905    & Experiment~\cite{ding2007structural} \\
&                         & 13.49    & DFT~\cite{kovci2008elasticity} \\
& \(B_0\) (GPa)          & 187.36   & Present work \\
&                         & 156.98 & Experiment~\cite{katahara1979pressure} \\

&                         & 195 & Experiment~\cite{ding2007structural} \\
&                         & 162 & Experiment~\cite{manghnani2000science} \\
&                         & 182 & DFT~\cite{kovci2008elasticity} \\
& \(dB_0/dP\)            & 3.8436   & Present work \\
&                         & 3.5 & Experiment~\cite{ding2007structural} \\
&                         & 3.5 & Experiment~\cite{manghnani2000science} \\
&                         & 3.75 & DFT~\cite{kovci2008elasticity} \\
\bottomrule
\end{tabular}
\end{table}

The complete RK interaction coefficients (\(L_0\), \(L_1\), and \(L_2\)) and fitting statistics are provided in the Supplementary Information (Table~S1 and S2). Fig.~\ref{fig:BM_parameters} shows the composition dependence of \(E_0\), \(V_0\), \(B_0\), and \(B_0^{\prime}\), together with the corresponding RK fits. The smooth trends and small fitting residuals further demonstrate that the isotropic equation-of-state parameters are well behaved across the entire composition range.

\begin{figure*}[h]
\centering
\includegraphics[width=0.95\textwidth]{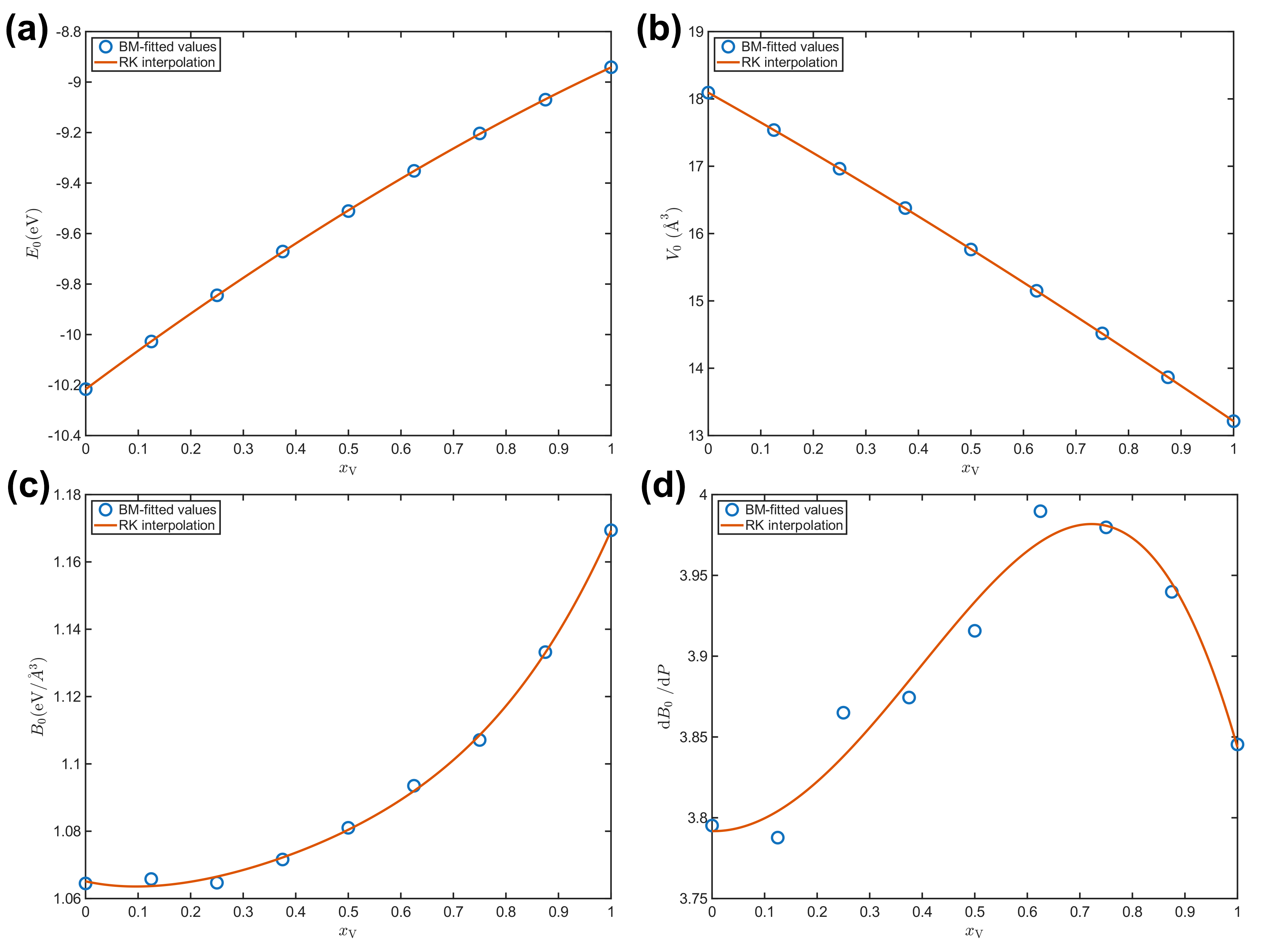}
\caption{Composition dependence of the fitted isotropic Birch--Murnaghan equation-of-state parameters for BCC Nb--V alloys. Blue circles denote the discrete parameter values obtained from independent Birch--Murnaghan fits to the DFT energy--volume data at sampled compositions, while solid lines represent the second-order Redlich--Kister interpolations. Panels show (a) equilibrium energy per atom \(E_0\), (b) equilibrium atomic volume \(V_0\), (c) bulk modulus \(B_0\), and (d) pressure derivative of the bulk modulus \(B_0^{\prime}\). The smooth trends and excellent agreement between discrete data and continuous fits confirm the robustness of the composition-dependent isotropic energetic model.}
\label{fig:BM_parameters}
\end{figure*}

We next examine the phase stability predicted by the isotropic model described above, in which two coexisting phases are constrained to share a single common atomic volume. Within this framework, the free energy of a homogeneous alloy at composition \(x\) and temperature \(T\) is written as
\begin{equation}
G^{1\phi}(x,T)=E_0(x)+\Delta G_{\mathrm{conf}}(x,T),
\end{equation}
where \(E_0(x)\) is obtained from the fitted Birch--Murnaghan model and \(\Delta G_{\mathrm{conf}}(x,T)\) is the ideal configurational entropy contribution. For a coherent two-phase state with phase compositions \(x_1\) and \(x_2\), phase fraction \(f=(x_2-x)/(x_2-x_1)\), and common constrained volume \(V_c\), the free energy is
\begin{equation}
G^{2\phi}(x_1,x_2,V_c;x,T)
=f\,G(x_1,V_c,T)+(1-f)\,G(x_2,V_c,T),
\end{equation}
with the equilibrium two-phase free energy obtained by minimizing with respect to \(V_c\). 

In practice, the numerical minimization of \(G^{2\phi}\) consistently collapses toward \(x_1=x_2=x\), indicating that no miscibility gap is predicted within the present fully coherent common-volume model. 

To understand this result, it is useful to compare the chemical driving force for decomposition with the corresponding coherent elastic penalty. Before imposing the common-volume constraint, the fitted \(E_0(x)\) curve is convex upward relative to the straight line connecting the pure end members, which means that the chemical contribution alone favors phase separation at low temperature. However, this driving force is extremely weak. For example, at \(x=0.5\), a small-amplitude decomposition into \(x_1=0.49\) and \(x_2=0.51\) gives
\begin{equation}
\Delta G_{\mathrm{chem}}
=\frac{1}{2}\left[E_0(x_1)+E_0(x_2)\right]-E_0(x)
=-2.743\times10^{-5}\ \mathrm{eV/atom},
\end{equation}
corresponding to only \(0.027\) meV/atom. At finite temperature, this chemical stabilization is further offset by the loss of configurational entropy, which is of the same order of magnitude for such a small composition split. 

Once the coherent constraint is imposed, both phases must deviate from their own equilibrium volumes and therefore incur a finite elastic penalty. For the same example \((x_1,x_2)=(0.49,0.51)\), the fitted equilibrium volumes differ as \(V_0(0.49)=15.818\) \AA\(^3\)/atom and \(V_0(0.51)=15.720\) \AA\(^3\)/atom. Minimization of the two-phase free energy with respect to the common volume gives an optimal constrained volume of \(V_c=15.769\) \AA\(^3\)/atom, at which the elastic contribution of the mixture is
\begin{equation}
\Delta G_{\mathrm{el}}^{\mathrm{mix}}(V_c)=8.253\times10^{-5}\ \mathrm{eV/atom}.
\end{equation}
This elastic penalty is approximately three times larger than the absolute value of chemical driving force, so that
\begin{equation}
\Delta G_{\mathrm{tot}}=\Delta G_{\mathrm{chem}}+\Delta G_{\mathrm{el}}^{\mathrm{mix}} > 0.
\end{equation}
Therefore, even this infinitesimal decomposition is thermodynamically unfavorable once the fully coherent common-volume constraint is enforced.

The present results show that phase separation under a fully isotropic coherent constraint carries a large elastic penalty. Because both phases are forced to share the same common volume, each phase deviates from its own equilibrium lattice parameter, and the associated strain energy exceeds the chemical driving force for decomposition. As a result, even at \(0\) K, infinitesimal composition fluctuations are not thermodynamically favored. The fully isotropic common-volume constraint therefore appears too restrictive for the Nb--V system.

This may seem inconsistent with experimental reports describing two BCC phases decomposition with nearly identical lattice parameters from XRD measurements~\cite{gao2018experimental}. However, small anisotropic distortions can be difficult to resolve experimentally when the deviation from cubic symmetry may be weak. Thus, an apparently ``two BCC phases with the same lattice constant'' interpretation does not necessarily imply a truly isotropic coherent state. This suggests that models allowing anisotropic relaxation, such as a biaxial coherent constraint, would be more realistic than the present isotropic model. 

\subsection{Composition-dependent energetic and elastic properties of biaxial model}
Since the fully isotropic common-volume constraint was found to be overly restrictive, we next consider the biaxial coherent model, in which the lattice parameters parallel to the interface remain continuous while relaxation normal to the interface is allowed.

With the settings introduced in Sec.~2.4, a representative fitted energy surface is shown in Fig.~2 for Nb$_{0.444}$V$_{0.556}$. The minimum-energy region forms a tilted valley in the $(f_{\perp},f_{\parallel})$ space, demonstrating a clear coupling between in-plane and out-of-plane deformation due to positive Poisson's ratio. In addition, the curvature of the surface is not perfectly symmetric with respect to tensile and compressive strain, reflecting the importance of anharmonic contributions beyond the harmonic elastic limit. This energy surface indicates that the expansion to the third order in Eq.~\ref{eq:biaxialexpand} is necessary, including quadratic terms, cubic terms, and mixed third-order terms.

\begin{figure}[h]
\centering
\includegraphics[width=0.8\textwidth]{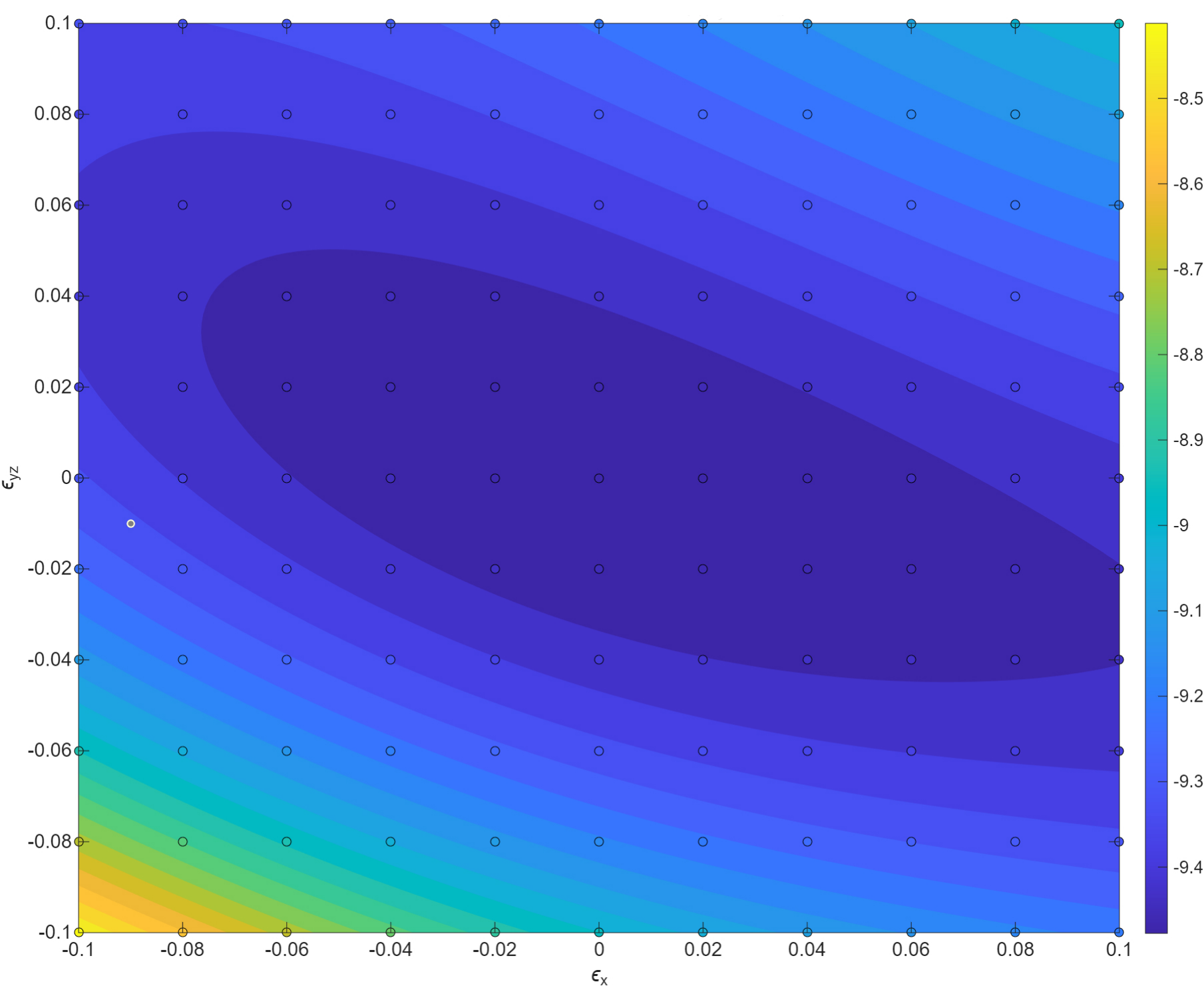}
\caption{Energy as a function of biaxial strain for Nb$_{0.444}$V$_{0.556}$. Symbols denote DFT data, and the surface corresponds to the fitted third-order model.}
\label{fig:example_energy_surface}
\end{figure}

For concentration of V from 0 to 1, the fitted coefficients, including $E_0$ and the strain-expansion parameters $A_{ij}$, vary smoothly with composition and were further represented using second-order Redlich--Kister expansions to provide a continuous energetic description across the full Nb--V composition range. Complete fitting coefficients, interpolation parameters, and fitting statistics are provided in the Supplementary Information (Fig.~S1, Table~S3 and Table~S4).

The resulting composition-dependent biaxial free-energy model is used in the following section to evaluate coherent phase stability, miscibility gaps, and spinodal behavior in the Nb--V system.

\subsection{Miscibility gap and spinodal with coherent elastic contribution}

To evaluate phase separation under coherent elastic constraint, we first determine the equilibrium lattice parameter at each discrete composition from the relaxed DFT structures, denoted by \(d_0(x)\). These values are then represented as a continuous function of composition using a second-order RK expansion. The temperature dependence is subsequently incorporated through the linear thermal expansion coefficient in Eq.~(24), yielding \(d_0(x,T)\). This \(d_0(x,T)\) serves as the strain-free reference lattice parameter for the corresponding homogeneous phase. For a phase constrained to adopt an in-plane lattice parameter \(d_{\parallel}\) and an out-of-plane lattice parameter \(d_{\perp}\), the corresponding stretch ratios are defined as
\begin{equation}
    \lambda_{\parallel}=\frac{d_{\parallel}}{d_0(x,T)}, \qquad
    \lambda_{\perp}=\frac{d_{\perp}}{d_0(x,T)}
\end{equation}

These stretch ratios are then converted to the finite-strain variables used in the elastic energy model.

The total Gibbs free energy of the coherent two-phase state at overall composition \(x\) is obtained by minimizing over phase compositions and lattice parameters:
\begin{equation}
\begin{aligned}
        G^{2\phi}(x,T)=
\min_{x_1,x_2,d_{\parallel},d_{\perp,1},d_{\perp,2}}
\Bigg\{
&k_1\,G(x_1,d_{\parallel},d_{\perp,1},T)\\
&+k_2\,G(x_2,d_{\parallel},d_{\perp,2},T)
\Bigg\}
\end{aligned}
\end{equation}

where $k_1$ and $k_2$ are the phase fractions
\begin{equation}
   k_1=\frac{x_2-X}{x_2-x_1}, \qquad
   k_2=\frac{X-x_1}{x_2-x_1}
\end{equation}

Because the two phases are coherent only in the interfacial in-plane directions, they share the same \(d_{\parallel}\), whereas their out-of-plane lattice parameters are independent. The above minimization can therefore be separated as
\begin{equation}
\begin{aligned}
G^{2\phi}(x,T)
= \min_{x_1,x_2,d_{\parallel}}
\Bigg\{
&k_1 \min_{d_{\perp,1}} G(x_1,d_{\parallel},d_{\perp,1},T) \\
&+ k_2 \min_{d_{\perp,2}} G(x_2,d_{\parallel},d_{\perp,2},T)
\Bigg\}
\end{aligned}
\end{equation}

Since the elastic contribution is quadratic in the out-of-plane degree of freedom for fixed \(x\), \(T\), and \(d_{\parallel}\), the minimization with respect to \(d_{\perp}\) can be carried out analytically.

The coherent binodal is determined by directly comparing the free energies of homogeneous and decomposed states on a discrete \((x,T)\) grid. Temperatures from 0 to 1800 K are sampled in intervals of 25 K, and compositions from 0 to 1 are sampled in intervals of 0.01. At each grid point, the single-phase Gibbs free energy is evaluated directly for the BCC solid solution. For the two-phase state, \(x_1\), \(x_2\), and \(d_{\parallel}\) are optimized under the constraint \(x_1<x<x_2\) to obtain the lowest coherent two-phase free energy. A point is identified as belonging to the two-phase region when the optimized decomposed state is lower in free energy than the single-phase state; otherwise it is classified as single phase. In practice, cases where the optimization collapses to \(x_1\approx x \approx x_2\) are also treated as single-phase states.

The coherent spinodal is estimated by testing the stability of the homogeneous phase against an infinitesimal composition modulation. For a given grid point \((x,T)\), we introduce a small composition perturbation \(\delta=0.001\) and compare the single-phase free energy \(G(x,T)\) with the coherent two-phase free energy constructed from two nearby compositions,
\begin{equation}
    \min_{d_{\parallel},d_{\perp,1},d_{\perp,2}}
G(x-\delta,x+\delta,d_{\parallel},d_{\perp,1},d_{\perp,2},T)
\end{equation}

If this infinitesimally decomposed state is lower in free energy than the homogeneous state, the alloy is regarded as unstable against small-amplitude decomposition and is classified inside the spinodal region. Otherwise, the homogeneous phase remains locally stable.

\begin{figure}[h]
\centering
\includegraphics[width=0.72\textwidth]{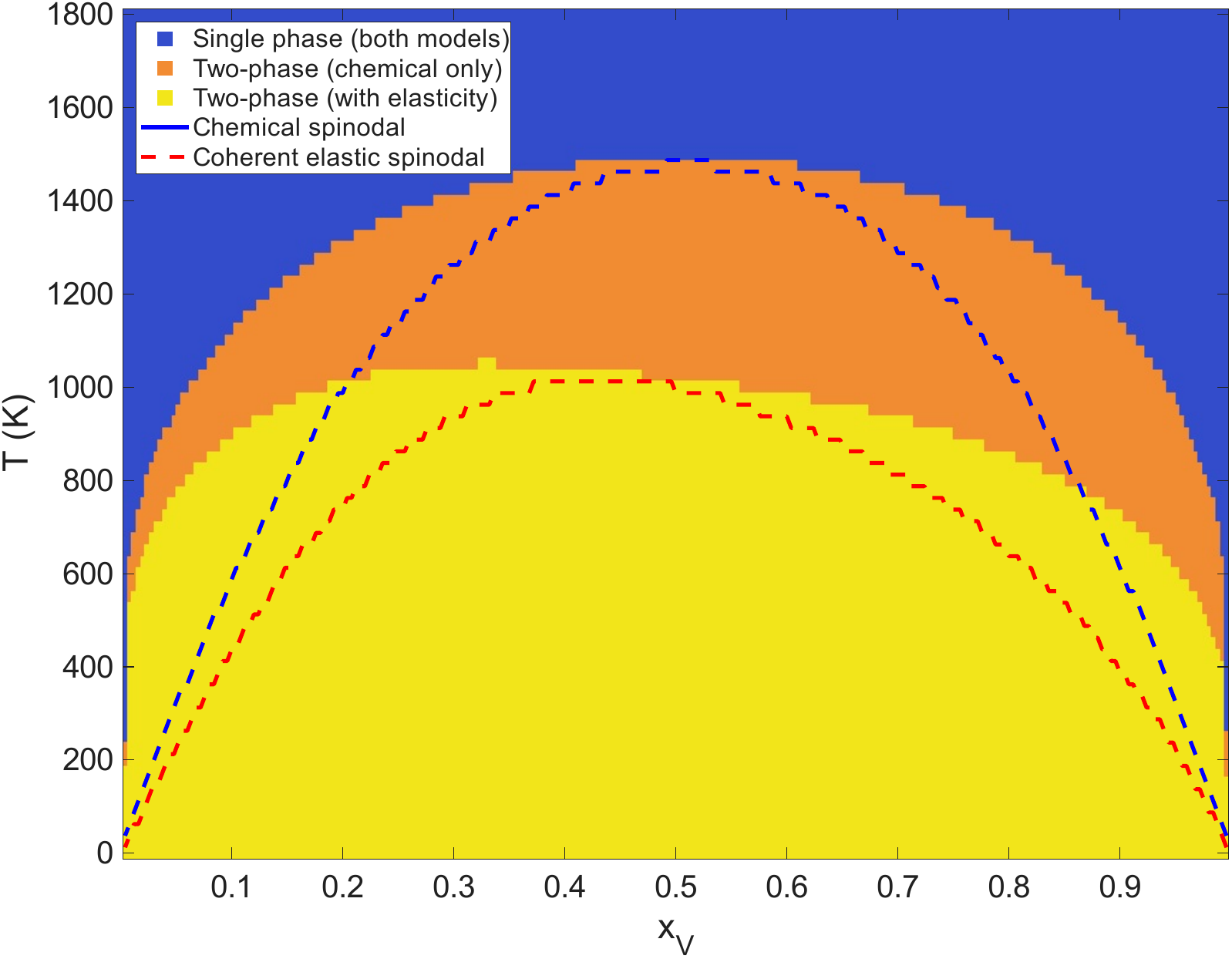}
\caption{Comparison of phase stability predicted by the chemical-only model and by the coherent-elastic model. Yellow: both models predict phase separation; orange: only the chemical-only model predicts two-phase equilibrium; blue: single-phase region. Dashed curves denote spinodal boundaries. Coherent elastic effects significantly reduce both the miscibility gap and the spinodal.}
\label{fig:miscibility_spinodal}
\end{figure}

Fig.~\ref{fig:miscibility_spinodal} compares the phase stability map obtained from the conventional chemical free energy alone and from the present model including coherent elastic contribution. In both descriptions, the outer boundary represents the binodal, while the dashed curves denote the spinodal limits. The chemical-only model predicts a broad miscibility gap with a critical temperature near 1475~K. Once the coherent elastic contribution is included, both the miscibility gap and the spinodal shrink substantially, and the critical temperature is reduced to approximately 1050~K, bringing the prediction into much closer agreement with the experimental value~\cite{gao2018experimental}. This reflects the energetic penalty associated with maintaining lattice coherence between decomposed phases, which stabilizes the homogeneous solid solution over a wider composition and temperature range.

This reduction is comparable in magnitude to other stabilizing mechanisms previously discussed in the literature, such as SRO, lattice distortion, and phonon entropy. At the same time, the large variation among previous predictions also indicates that the quantitative influence of each contribution remains strongly model dependent. For example, vibrational effects were found to be minor in the work of Wang \emph{et al.}, but substantial in Ravi \emph{et al.}, where phonons lowered the critical temperature by nearly 300 K.

Importantly, none of the previous studies explicitly considered the coherent elastic penalty between two coexisting decomposed phases under lattice-compatibility constraints. Although Ravi \emph{et al.} introduced a constituent strain term within the CE Hamiltonian, this contribution acts primarily as a configuration-dependent long-range interaction in the single-phase alloy and was found to have only a minor effect on the Nb–V phase boundary~\cite{ravi2012first}. By contrast, the present work directly evaluates the elastic penalty associated with maintaining coherency between two separated phases with different equilibrium lattice parameters. This penalty is shown to be sufficiently large to strongly suppress phase separation and stabilize the homogeneous solid solution.

Both theory and experiment suggest that many factors favor the single-phase BCC solution and reduce the miscibility gap in Nb–V alloys, including SRO, lattice distortion, phonon entropy, and the coherent elastic effect identified here. In practice, however, these contributions are highly coupled and difficult to treat simultaneously. The degree of SRO and the vibrational free energy may depend on the strain state of coherently constrained two-phase microstructures, while the equilibrium coherent strain may be altered once SRO is included. Constructing a fully self-consistent model that captures all of these effects remains a major challenge.

Additional complications arise in real materials, where interfaces may relax through dislocations, grain boundaries, or other defects that partially release coherency stresses and modify the observed phase boundaries. The experimentally reported miscibility gap also remains noticeably narrower than most theoretical predictions, despite comparable critical temperatures. This likely reflects the extremely sluggish diffusion kinetics in Nb–V alloys, which makes equilibrium decomposition difficult to achieve; Gao \emph{et al.} reported annealing treatments as long as 70 days at $600~^\circ\mathrm{C}$, and previous studies noted that slow diffusion could obscure phase separation experimentally. Such kinetic and microstructural factors likely contribute to the remaining discrepancies between theory and experiment.

The main contribution of the present work is therefore not to claim a definitive phase diagram for Nb–V, but to identify coherent elastic energy as a previously neglected thermodynamic factor, demonstrate that its magnitude is comparable to other stabilization mechanisms, and show that incorporating this effect systematically shifts predictions toward experiment. More broadly, coherent elasticity should be considered explicitly whenever phase separation occurs between solids with appreciable lattice mismatch.

\begin{figure}[H]
\centering
\includegraphics[width=0.95\textwidth]{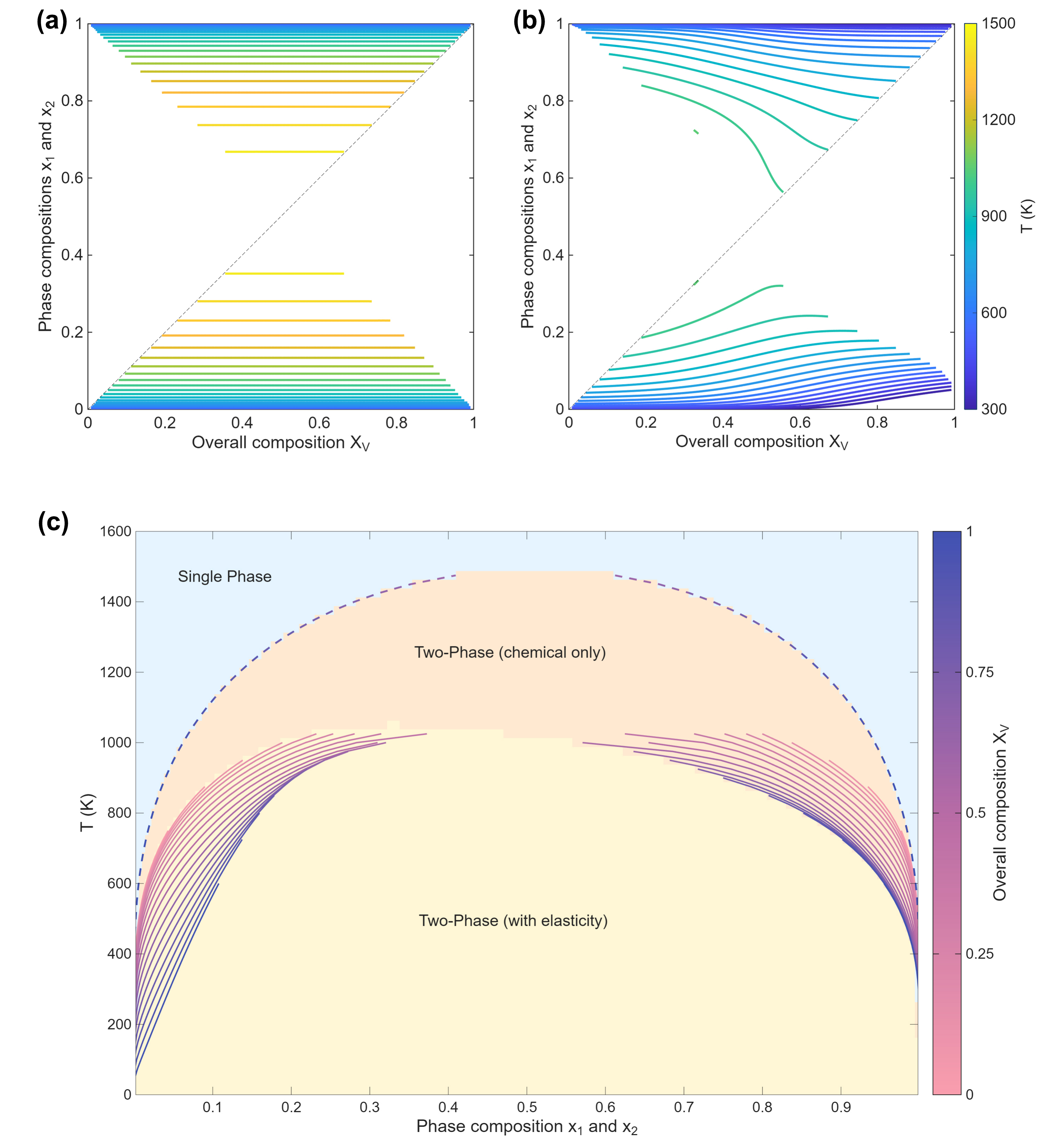}
\caption{
Equilibrium decomposition compositions under chemical-only and coherent-elastic descriptions.
(a) Chemical-only model: all overall compositions \(X\) within the miscibility gap decompose into the same pair \((x_1,x_2)\) at a given temperature, giving horizontal tie-lines consistent with the lever rule.
(b) Coherent-elastic model: the decomposition compositions become functions of \(X\). Increasing \(X_{\mathrm{V}}\) at fixed temperature shifts \(x_1\) upward and \(x_2\) downward, indicating weaker decomposition on the V-rich side due to the larger coherent elastic penalty.
(c) Temperature dependence of decomposition trajectories for multiple \(X_{\mathrm{V}}\). Under coherent constraint, each overall composition follows a distinct path. The envelope of these trajectories reconstructs the miscibility-gap boundary, showing that the phase boundary represents the stability limit of composition-dependent decomposed states rather than unique coexistence compositions.
}
\label{fig:x1x2}
\end{figure}

Fig.~\ref{fig:x1x2} further presents the relationship between the overall composition \(X\) and the resulting equilibrium phase compositions \(x_1\) and \(x_2\) at different temperatures, for both the chemical-only and coherent-elastic models. In the chemical-only case in Fig.~\ref{fig:x1x2}(a), for a given temperature, all overall compositions \(X\) within the miscibility gap decompose into the same pair of equilibrium compositions \(x_1\) and \(x_2\). This corresponds to the conventional picture of phase separation governed by tie-lines and the lever rule: the coexistence compositions depend only on temperature and are independent of the overall composition. As a result, the curves in Fig.~\ref{fig:x1x2}(a) are horizontal for each temperature.

\newpage
In contrast, when coherent elastic effects are included, as shown in Fig.~\ref{fig:x1x2}(b), the decomposition behavior changes qualitatively. In addition to the previously observed reduction of the critical temperature and narrowing of the miscibility gap, the equilibrium phase compositions \(x_1\) and \(x_2\) become explicit functions of the overall composition \(X\). Specifically, at fixed temperature, increasing the overall composition \(X_{\mathrm{V}}\) leads to an increase in \(x_1\) and a decrease in \(x_2\), indicating that the degree of decomposition is progressively suppressed on the V-rich side. As \(X\) approaches either boundary of the miscibility gap, one phase fraction vanishes and the corresponding phase composition necessarily approaches \(X\). The intersection of the curves with the diagonal \(x=X\) therefore remains valid under coherent elasticity, except for minor numerical artifacts near the critical temperature in Fig.~\ref{fig:x1x2}(b).

Fig.~\ref{fig:x1x2}(c) further illustrates the temperature dependence of the decomposition behavior for different overall compositions \(X_{\mathrm{V}}\), by plotting the \(x_1\)--\(T\) and \(x_2\)--\(T\) trajectories for \(X_{\mathrm{V}}\) ranging from 0.05 to 0.95. In the chemical-only case, these trajectories collapse onto the conventional binodal boundaries, and horizontal tie-lines can be drawn at each temperature to connect the corresponding \(x_1\) and \(x_2\) values, consistent with the standard phase-diagram interpretation. However, in the presence of coherent elastic energy, although a well-defined miscibility gap and single-phase boundary still exist, the decomposition compositions \(x_1\) and \(x_2\) no longer lie on the binodal boundary, nor are they independent of \(X\). Instead, each overall composition follows a distinct decomposition trajectory. Remarkably, the envelope formed by these \(x_1(T)\) and \(x_2(T)\) curves across different \(X_{\mathrm{V}}\) reconstructs the boundary of the miscibility gap. This can be understood from the relationship between Fig.~\ref{fig:x1x2}(b) and Fig.~\ref{fig:x1x2}(c). At a given temperature, the allowed range of overall compositions \(X\) within the miscibility gap corresponds to the horizontal extent of the curves in Fig.~\ref{fig:x1x2}(b). The minimum value of \(X\) is given by the leftmost intersection of the curves with the diagonal line \(x_1=X\), while the maximum value is given by the rightmost intersection with \(x_2=X\). Because \(x_1\) increases monotonically with \(X\) and \(x_2\) decreases monotonically with \(X\) in the present system, as shown in Fig.~\ref{fig:x1x2}(b), the accessible range of \(X\) at each temperature is determined by the minimum of \(x_1\) and the maximum of \(x_2\). When mapped onto the \(x\)--\(T\) plane in Fig.~\ref{fig:x1x2}(c), this translates into the outer envelope of the \(x_1(T)\) curves and the inner envelope of the \(x_2(T)\) curves, which together define the miscibility-gap boundary.

\begin{figure}[t]
\centering
\includegraphics[width=0.95\textwidth]{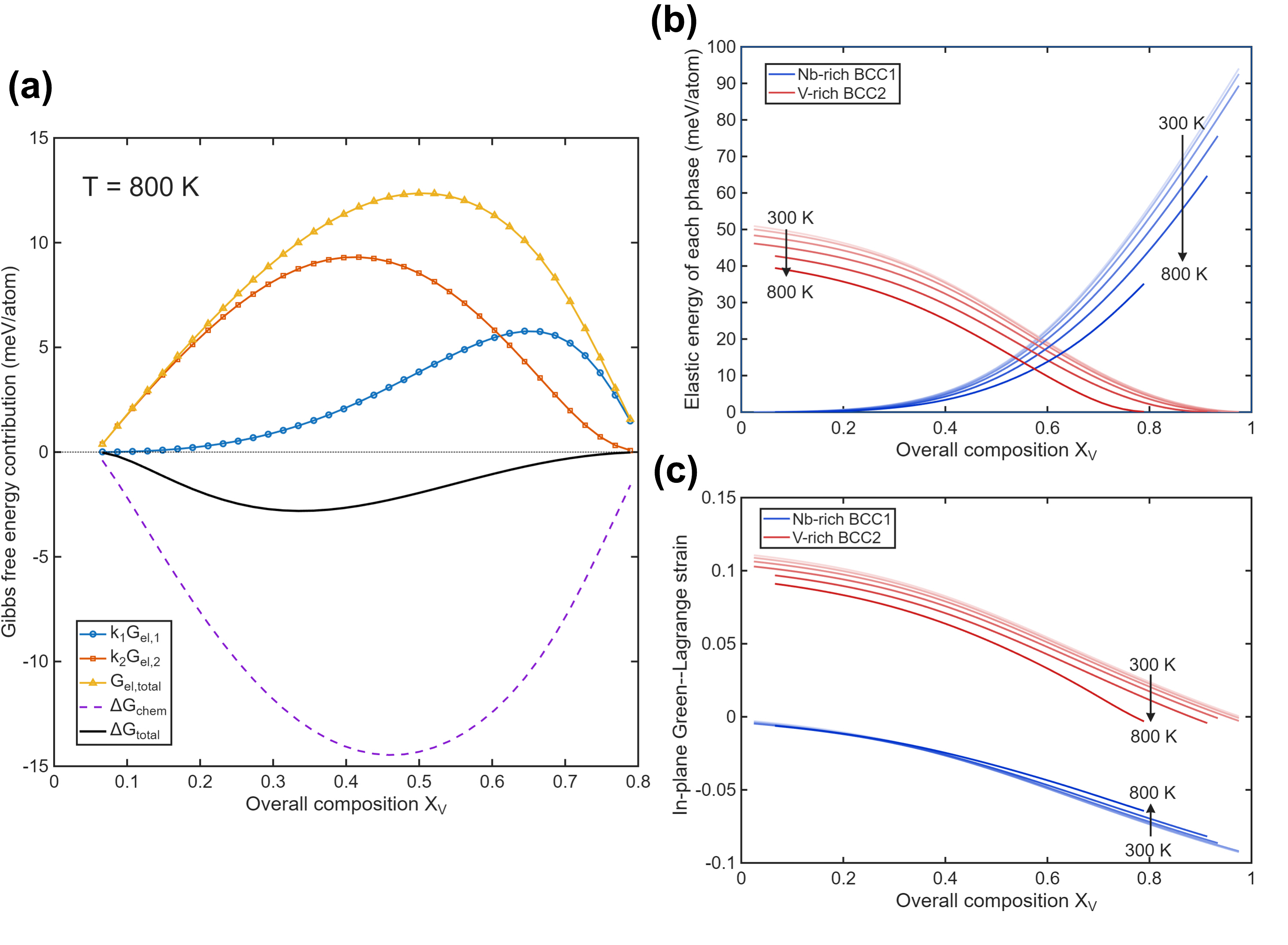}
\caption{
Analysis of the elastic origin of the composition-dependent phase behavior under coherent constraint.
(a) Free-energy decomposition at $T=800$ K, showing the chemical driving force $\Delta G_{\mathrm{chem}}$, the elastic contribution of each phase, and the resulting total $\Delta G_{\mathrm{total}}$. (b) Phase-resolved elastic energies per atom as a function of overall composition for different temperatures. The variation with temperature is indicated by the color intensity. (c) Phase-resolved in-plane Green--Lagrange strain $f_{\parallel}$ as a function of overall composition for different temperatures. The variation with temperature is indicated by the color intensity. The Nb-rich phase is under compression ($f_{\parallel}<0$), while the V-rich phase is under tension ($f_{\parallel}>0$). Near the phase boundaries, the majority phase remains close to its equilibrium state, whereas the minority phase accommodates most of the deformation. Notably, the tensile strain in the V-rich phase at low $X_{\mathrm{V}}$ is larger in magnitude than the compressive strain in the Nb-rich phase at high $X_{\mathrm{V}}$.
}
\label{fig:elastic_mechanism}
\end{figure}

We note that the above conclusions are closely associated with the monotonic dependence of the equilibrium phase compositions on the overall composition. To better understand the origin of this monotonic behavior, we further analyze the phase-resolved elastic contributions and the corresponding strain states, as shown in Fig.~\ref{fig:elastic_mechanism}. 

In Fig.~\ref{fig:elastic_mechanism}(a), the total elastic contribution, $G_{\mathrm{el,total}} = f_1 G_{\mathrm{el},1} + f_2 G_{\mathrm{el},2}$, acts in opposition to the chemical driving force for phase separation. While the chemical free energy favors decomposition, the elastic energy penalizes the compositional mismatch between the two phases. As a result, the net driving force for phase separation (black curve) remains negative but is significantly reduced compared to the chemical-only case, indicating that coherent elasticity strongly suppresses phase separation. A key feature revealed in Fig.~\ref{fig:elastic_mechanism}(a) is that the elastic energy is not uniformly distributed between the two phases. Instead, it is dominated by the minority phase which deviates most from its equilibrium lattice parameter under the coherent constraint. At small overall composition $X_V$, the optimized in-plane lattice parameter is close to that of the Nb-rich phase, as shown in Fig.~\ref{fig:elastic_mechanism}(c). Consequently, the V-rich phase accommodates most of the lattice mismatch and dominates the elastic contribution, even though its phase fraction is small. Conversely, at large $X_V$, the Nb-rich phase becomes the one that is significantly distorted, and thus its elastic contribution dominates the total elastic energy despite its minority fraction.

Importantly, the elastic responses of the two phases are strongly asymmetric. As shown in Fig.~\ref{fig:elastic_mechanism}(b), the elastic energy of the Nb-rich phase increases much more rapidly than that of the V-rich phase. This asymmetry arises from two factors. First, there is an intrinsic difference between compression and tension: compressive strain is energetically more costly than tensile strain, as reflected in Fig.\ref{fig:example_energy_surface} and the positive $dB_0/dP$ in the BM model. Second, the V-rich compositions exhibit smaller elastic coefficients $A_{ij}$ (Table~S3), indicating a lower stiffness compared to Nb-rich compositions. Together, these effects make the V-rich phase more tolerant to tensile deformation, while the Nb-rich phase exhibits a much stronger resistance to compressive strain.

This behavior is further illustrated in Fig.~\ref{fig:elastic_mechanism}(c), where the Nb-rich phase is under compressive strain (negative) and the V-rich phase is under tensile strain (positive). Near the phase boundaries, the majority phase remains close to its equilibrium state, while the minority phase accommodates most of the deformation. Notably, the magnitude of tensile strain in the V-rich phase exceeds the corresponding compressive strain in the Nb-rich phase at the opposite ends, indicating that biaxial tensile deformation of V-rich phase is more easily accommodated than biaxial compression of Nb-rich phase.

As a consequence, with increasing $X_V$, the elastic penalty associated with compressing the Nb-rich phase grows rapidly and becomes dominant. To reduce this growing penalty, the system further suppresses the compositional mismatch between the two phases as $X_{\mathrm{V}}$ increases. This leads to an increase of $x_1$ and a decrease of $x_2$ --- a progressive contraction of the two-phase region toward the overall composition observed in Fig.~\ref{fig:x1x2}(b), which underlies the envelope construction that defines the phase boundary in Fig.~\ref{fig:x1x2}(c).

Given that the envelope construction of the phase boundary relies on the monotonic dependence of the equilibrium phase compositions on the overall composition, this behavior is not expected to be universal. The monotonicity analysis above is based on the composition dependence of the elastic properties in the specific Nb--V system, which is representative of many alloy systems with relatively smooth composition dependence. In principle, for systems in which the elastic energy or equilibrium lattice parameter exhibits strongly nonlinear dependence on composition, the functions \(x_1(X,T)\) and \(x_2(X,T)\) may become non-monotonic. In such cases, the resulting miscibility-gap boundary in the \(X\)--\(T\) space could deviate from the conventional convex shape, and more complex topologies may arise. However, this would require sufficiently strong nonlinearity in the elastic contribution or a relatively low chemical driving force of phase decomposition.

\section{Conclusions}

In this work, we develop a thermodynamic framework to examine the influence of coherent elastic strain on phase separation in the BCC Nb--V alloy system. By combining composition-dependent chemical free energies from first-principles calculations with coherent elastic (both isotropic and biaxial strain) energy contributions, we construct a coherent-elastic phase diagram and compare it with the conventional chemical-only description. The results show that coherent elastic strain helps stabilize the single BCC phase and strongly suppresses phase separation in BCC Nb--V. Relative to the chemical-only model, the miscibility gap is substantially narrowed and the critical temperature is reduced, bringing the prediction into much closer agreement with available experimental observations. This demonstrates that coherent strain energy provides an important stabilization of the single-phase solid solution when lattice mismatch between decomposed phases is appreciable. The present calculations further suggest that the decomposed Nb-rich and V-rich phases might adopt nearly tetragonal structures rather than perfectly cubic BCC lattices, sharing a common in-plane lattice parameter while relaxing differently along the perpendicular direction. Because the predicted distortion is small, such deviations may be difficult to resolve experimentally.

More importantly, we find that coherent elasticity qualitatively changes the conventional interpretation of phase equilibria. In the chemical-only picture, coexistence compositions depend only on temperature and are represented by horizontal tie-lines. Under coherent constraint, however, the equilibrium decomposition products depend on both temperature and the overall alloy composition. Consequently, the two-phase boundary no longer corresponds to unique temperature-dependent coexistence compositions, but instead represents the stability limit of composition-dependent coherent decomposed states.

This study shows that coherent elastic strain modifies not only the quantitative extent of a miscibility gap, but also the physical meaning of phase boundaries and coexistence compositions. These results provide a broader perspective for understanding phase separation in coherent alloys. Future work may proceed in three directions: developing more general descriptions of coherency beyond the present biaxial model; incorporating short-range order and phonon contributions, possibly as an initial approximation in which these effects are treated independently from coherency; and and extending the framework to other alloy systems to assess its generality, followed by applications to systems with larger lattice mismatch, where coherent elastic effects may be even more pronounced.

\section*{Acknowledgements}
The authors would like to acknowledge the support of the National Science Foundation through Grant No. 2119103 and the U.S. Department of Energy (DOE) ARPA-E ULTIMATE Program through Project DE-AR0001427. RA acknowledges support from the Army Research Laboratory and was accomplished under Cooperative Agreement Number W911NF-22-2-0106. Calculations were carried out at the Texas A\&M High-Performance Research Computing (HPRC) Facility.

\bibliographystyle{elsarticle-num}
\bibliography{ref}

\end{document}